# Amplified Photocurrent in Heterojunctions comprising Nano-rippled Zinc Oxide and Perovskite-inspired $Cs_3Cu_2I_5$

Si Hyeok Yang[a+], Lim Kyung Oh[b+], Na Young Lee[a+], Dong Ho Lee[b+], Sang Min Choi[a], Bowon Oh[a], Yun Ji Park[a], Yunji Cho[c], Jaesel Ryu[d], Hongki Kim[c*], Sang-Hyun Chin[e*], Yeonjin Yi[e*], Myungkwan Song[f*], Han Seul Kim[bg*], and Jin Woo Choi[a*]

[a] Department of Data Information and Physics, Kongju National University, Gongju 32588, Republic of Korea. Email: jinwoo.choi@kongju.ac.kr

[b] Department of Advanced Materials Engineering, Chungbuk National University, Cheongju 28644, Republic of Korea. Email: hanseul.kim@chungbuk.ac.kr

[c] Department of Chemistry, Kongju National University, Gongju 32588, Republic of Korea. Email: hongkikim@kongju.ac.kr

[d] Department of Atmospheric Sciences, Kongju National University, Gongju 32588, Republic of Korea.

[e] Department of Physics, Yonsei University, Yonsei-ro 50, Seoul, 03722, Republic of Korea. Email: sanghyunchin@yonsei.ac.kr; yeonjin@yonsei.ac.kr

[f] Department of Energy & Electronic Materials, Korea Institute of Materials Science (KIMS), Changwon, 51508, Republic of Korea. Email: smk1017@kims.re.kr

[g] Department of Urban, Energy, and Environmental Engineering, Chungbuk National University, Cheongju 28644, Republic of Korea.

[+] These authors contributed equally to this work.

S. H. Yang: https://orcid.org/0009-0002-1617-6898

H. Kim: https://orcid.org/0000-0002-5786-7030

S. H. Chin: https://orcid.org/0000-0002-0963-4777

Y. Yi: https://orcid.org/0000-0003-4944-8319

M. Song: https://orcid.org/0000-0001-7935-5303

H. S. Kim: https://orcid.org/0000-0003-0731-6705

J. W. Choi: https://orcid.org/0000-0001-6578-8132

**Abstract**

Molecular zero-dimensional (0D) halide perovskite-inspired cesium copper iodide ($Cs_3Cu_2I_5$) is a highly promising candidate for optoelectronic applications due to their low toxicity, high stability, and intense blue emission. However, their intrinsically poor electrical conductivity, stemming from isolated conductive copper iodide tetrahedra by cesium atoms, severely limits charge transport which poses a critical challenge for optoelectronic applications. In this study, we propose a novel strategy to overcome this limitation by utilizing precisely optimized zinc oxide nanoripple structures within a lateral $Cs_3Cu_2I_5$ photodetector (PD) architecture featuring interdigitated electrodes (IDEs). The ZnO nanoripple was systematically tuned to improve the percolation paths, providing efficient routes for photogenerated carriers to migrate to the IDEs. Consequently, the optimized heterojunctions comprising $Cs_3Cu_2I_5$ and ZnO exhibited superior photocurrent compared to the pristine $Cs_3Cu_2I_5$ counterparts. This nanostructure-mediated charge transport engineering strategy for lateral structured PDs offers a new pathway for utilizing low-conductivity 0D materials for conventional optoelectronics, next-generation

Internet of Things sensor networks, and plausibly biosensing applications.

**Introduction**

The rapid advancement of next-generation optoelectronic devices and communication technologies necessitates the development of high-performance photodetectors (PDs) exhibiting high sensitivity and ultrafast response across a broad spectral range.[1] The structural design of PDs is a critical determinant of their performance, broadly categorized into vertical and lateral configurations based on the arrangement of the photoactive layer and electrodes. Traditional vertical structures place electrodes above and below the active layer. This device architecture limits the charge separation distance within the photoactive region and causes disadvantages for achieving fast response times and high external quantum efficiency. In contrast, lateral structures leverage interdigitated electrodes (IDEs) to shorten the in-plane charge transport channel, enabling high photoconductive gain via efficient in-plane carrier transport.[2,3] For materials with intrinsically low light absorption or short carrier lifetimes, utilization of lateral IDEs for efficient charge collection has emerged as a key strategy for high-performance device fabrication. Recently, the high performance of lead halide perovskites has driven significant progress in optoelectronics, yet their long-term toxicity and instability remain major hurdles for commercialization.[4] Following this trend, low-toxicity optoelectronic materials, especially perovskite analogues such as $Cs_3Cu_2I_5$ have attracted significant attention due to their exceptional luminescence.[4–17] However, a major technological challenge arises from the intrinsically poor conductivity of these 0D materials. This material, $Cs_3Cu_2I_5$, is classified as a molecular-level low dimensional material, meaning the dimensional reduction and subsequent isolation of the conductive units that occur at the molecular level, rather than just the nanoscale morphology.[18] This distinction is crucial: its poor conductivity is not merely a consequence of nanoscale size, but rather stems directly from the material's crystal structure, specifically the isolated conductive copper iodide tetrahedra separated by cesium atoms. In addition, $Cs_3Cu_2I_5$ films typically grow in an island-like manner, where the isolated crystallites further disrupt the charge percolation network, making lateral charge transport between neighboring grains inefficient. To overcome these intrinsic limitations, engineering the underlying substrate to introduce nanoscale percolation pathways provides a promising route to amplify the photocurrent response. In particular, the formation of nanoripple structures on ZnO surfaces can facilitate lateral carrier transport by providing continuous percolation paths between otherwise isolated $Cs_3Cu_2I_5$ islands, thereby enhancing charge extraction efficiency and overall device performance. Previous research has shown that integrating a ZnO thin film as an interfacial layer and forming a nanoripple structure on its surface effectively increases the active contact area and improves the charge transport path in optoelectronic devices.[19]

Herein, we aim to integrate these structural and material advantages to realize a high-performance lateral $Cs_3Cu_2I_5$-based photodetector. Specifically, we precisely control and optimize the spontaneously formed nanoripple structure on the ZnO thin film prior to the deposition of the $Cs_3Cu_2I_5$ active layer. This optimized ZnO nanoripple structure profoundly influences the crystallization and interfacial morphology of the $Cs_3Cu_2I_5$ film. Moreover, this underlying structure drastically increases the percolation paths which are the necessary routes for charge carriers to efficiently migrate to the IDEs, overcoming the inherent low conductivity of 0D materials. This strategy maximizes the charge collection efficiency at the $Cs_3Cu_2I_5$/ZnO interfaces, resulting in a significant improvement in the photoresponsivity and response speed of the final IDEs-based lateral device compared to conventional counterparts. Ultimately, the nanostructure-mediated charge transport engineering strategy presented in this work can be applied to fields demanding high sensitivity and fast response. Beyond basic optical detection in wearable and implantable devices, the lateral structure with "exposed optically active layer" offers an exciting opportunity by functionalizing bioreceptors directly onto them.[20]

## 2. Materials and Methods

***Fabrication of the ZnO*** Zinc acetate was dissolved in a mixed solvent of 2-methoxyethanol and ethanolamine at a volume ratio of 10:1 to prepare a 1.5 M solution. After stirring, the solution was filtered through a PTFE filter with a pore size of 0.45 μm to remove impurities.

***Fabrication of the $Cs_3Cu_2I_5$*** CsI and CuI were dissolved in DMF at a molar ratio of 3:2 to prepare the 2.0 M precursor solutions. The precursor was stirred at 60 °C for 24 h inside a glove box. After the stirring process, the solutions were filtered through PTFE filters with a pore size of 0.45 μm to remove impurities.

***Device Fabrication*** The device was fabricated on glass. The substrates were cleaned using sonication in deionized (DI) water, acetone, and IPA for 15 min. The ZnO film was prepared by spin-coating the precursor solution at 3000 rpm for 60 s, followed by thermal treatment at 125 °C for 10 min. The devices were deposited 100 nm-thick Ag with a mask as an electrode. Finally, the $Cs_3Cu_2I_5$ layer was spin-coated onto the ZnO film at 3000 rpm for 60 s in a glove box and subsequently annealed to complete the device fabrication.

***Assessment of Photodetectors*** All electrical measurements were conducted using the Keithley2601-B source meter. The devices were irradiated under monochromatic light with wavelengths ranging from 390 to 660 nm (Ossila), providing equal light intensity for each wavelength to evaluate the photo-responsive characteristics.

## 3. Results and Discussion

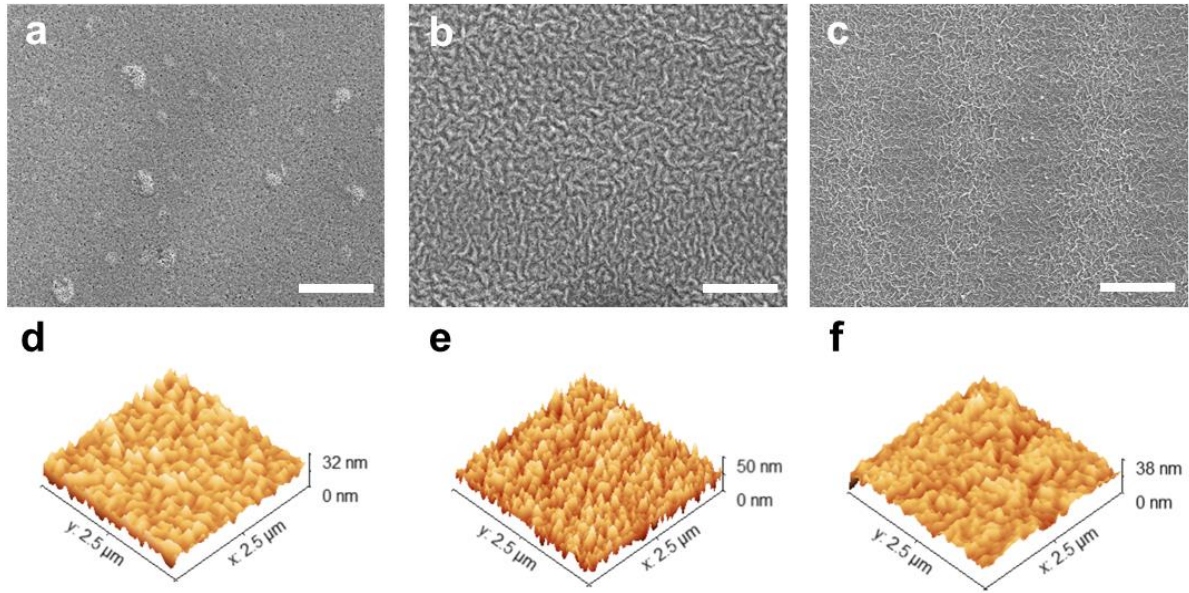

**Fig. 1.** Crystal structure of Scanning electron microscope images of ZnO films annealed at **a** 100 °C, **b** 125 °C, and **c** 150 °C (scale bar: 50 µm). Corresponding laser scanning microscope images of ZnO films annealed at **d** 100 °C, **e** 125 °C, and **f** 150 °C.

ZnO nanoripples are realized onto the substrates to establish percolation path, which subsequently aids in forming continuous charge transport channels between the $Cs_3Cu_2I_5$ islands. Therefore, **Fig. 1** details the process of precisely controlling and optimizing the ZnO nanoripple structure via annealing temperature to achieve this specific goal. **Fig. 1a-c** display the scanning electron microscope images of ZnO films annealed at 100 °C, 125 °C, and 150 °C, respectively. The ZnO annealed at 100 °C (**Fig. 1a**) exhibits a relatively flat and agglomerated particle morphology. In contrast, the ZnO annealed at 125 °C (**Fig. 1b**) clearly shows a uniform and densely distributed, fine nanoripple structure across the entire surface. This structure represents the optimized form, intended to serve as effective percolation paths for charge carrier movement. This morphological change is further clarified in three dimensions by the laser scanning microscope images **Fig. 1d-f**, which visualize the surface roughness and the height of the features. Compared to the 100 °C (**Fig. 1d**) and 150 °C (**Fig. 1f**) samples, the 125 °C (**Fig. 1b**) forms the largest height variation, evidenced by a maximum feature height reaching 50 nm on the z-axis scale. This well-developed nanoripple structure suggests the most ideal interfacial morphology for maximizing charge collection and transport efficiency by providing continuous charge transfer channels between the subsequent $Cs_3Cu_2I_5$ islands.

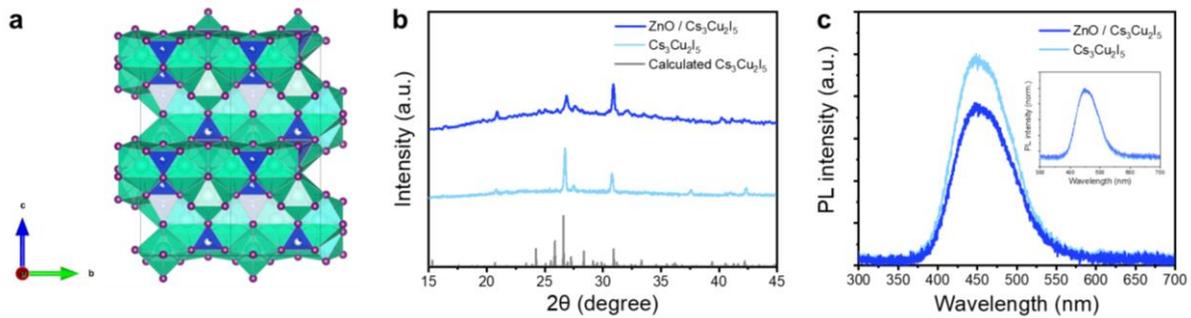

**Fig. 2. a** Crystal structure of $Cs_3Cu_2I_5$ (CCDC-2382223). **b** X-ray diffraction pattern of $Cs_3Cu_2I_5$ and ZnO/$Cs_3Cu_2I_5$ heterostructures. **c** Photoluminescence (PL) spectra of $Cs_3Cu_2I_5$ and ZnO/$Cs_3Cu_2I_5$ heterostructures (inset: normalized PL spectra of $Cs_3Cu_2I_5$ and ZnO/$Cs_3Cu_2I_5$ heterostructures).

**Fig. 2a** illustrates the crystal structure (CCDC-2382223) of the photoactive layer, molecular zero-dimensional (0D) $Cs_3Cu_2I_5$. The structure consists of $[Cu_2I_5]^{3-}$ and $Cs^+$, where the conductive copper iodide tetrahedra are completely isolated (Green: Cs, Blue: Cu, Purple: I).

This isolated configuration directly accounts for the material's intrinsically poor electrical conductivity. The following panels confirm that integrating the ZnO layer preserves this crystal structure while simultaneously enhancing charge extraction. **Fig. 2b** presents the X-ray diffraction (XRD) patterns of the pristine $Cs_3Cu_2I_5$ thin film and the $ZnO/Cs_3Cu_2I_5$ heterojunction. Clear characteristic peaks are observed in both samples, confirming the crystalline nature of the $Cs_3Cu_2I_5$ active layer. The main $Cs_3Cu_2I_5$ peaks remain well preserved in the XRD pattern of the heterojunction, indicating that the active layer was successfully formed on the nano-rippled ZnO surface and that the interfacial process did not significantly disturb the inherent crystal structure of $Cs_3Cu_2I_5$. **Fig. 2c** compares the photoluminescence (PL) spectra of the two structures. Both samples exhibit the characteristic blue emission of $Cs_3Cu_2I_5$ and maintain similar peak positions. A notable observation is the relative decrease in PL intensity for the $ZnO/ Cs_3Cu_2I_5$ heterojunction compared to the pristine $Cs_3Cu_2I_5$ film. This result strongly suggests that efficient charge separation and transport occur at the $ZnO/Cs_3Cu_2I_5$ interface, effectively competing with radiative recombination. The reduced PL intensity thus serves as key optical evidence supporting the successful enhancement of charge extraction efficiency, which is the main objective for improving photodetector performance in this study.

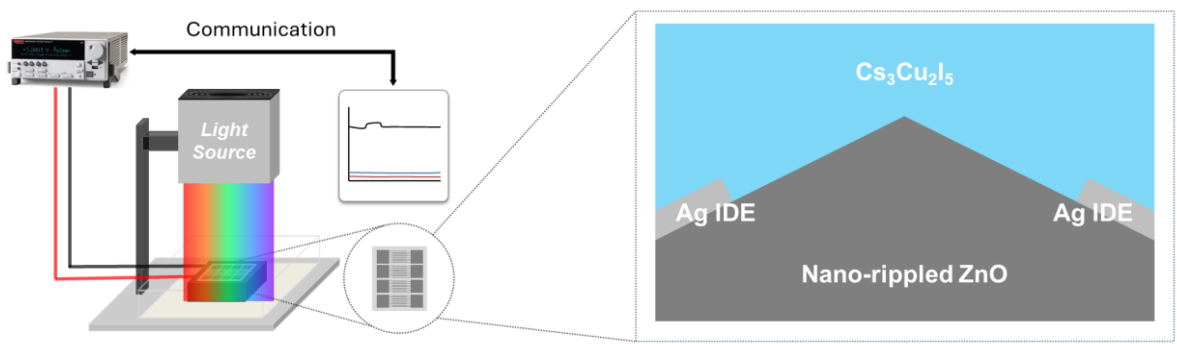

**Fig. 3.** Schematic illustration of silver IDE-based $ZnO/Cs_3Cu_2I_5$ heterojunctions and measurement set-up.

**Fig. 3** provides a schematic illustration of the final Ag interdigitated electrode (IDE)-based ZnO/ $Cs_3Cu_2I_5$ heterojunction photodetector and the corresponding measurement setup. The device utilizes a lateral structure that is crucial for amplifying the photocurrent by efficiently shortening the in-plane charge transport channel, especially for low-conductivity materials. The device stack consists of the nano-rippled ZnO layer deposited on the substrate, followed by the patterning of Ag IDE. The photoactive $Cs_3Cu_2I_5$ layer is then coated over the ZnO surface and the Ag IDE. The magnified view clearly highlights the role of the underlying nano-rippled ZnO layer, which is specifically optimized to provide continuous percolation paths for efficient lateral carrier transport at the $Cs_3Cu_2I_5$ / ZnO interface, thereby addressing the intrinsic poor conductivity of the $Cs_3Cu_2I_5$. The electrical characterization is performed by connecting the Ag IDE to an electric source meter, which monitors the current response while the device is irradiated with monochromatic light from the "Light Source". This setup is designed to assess the photo-responsive characteristics of the optimized heterojunction.

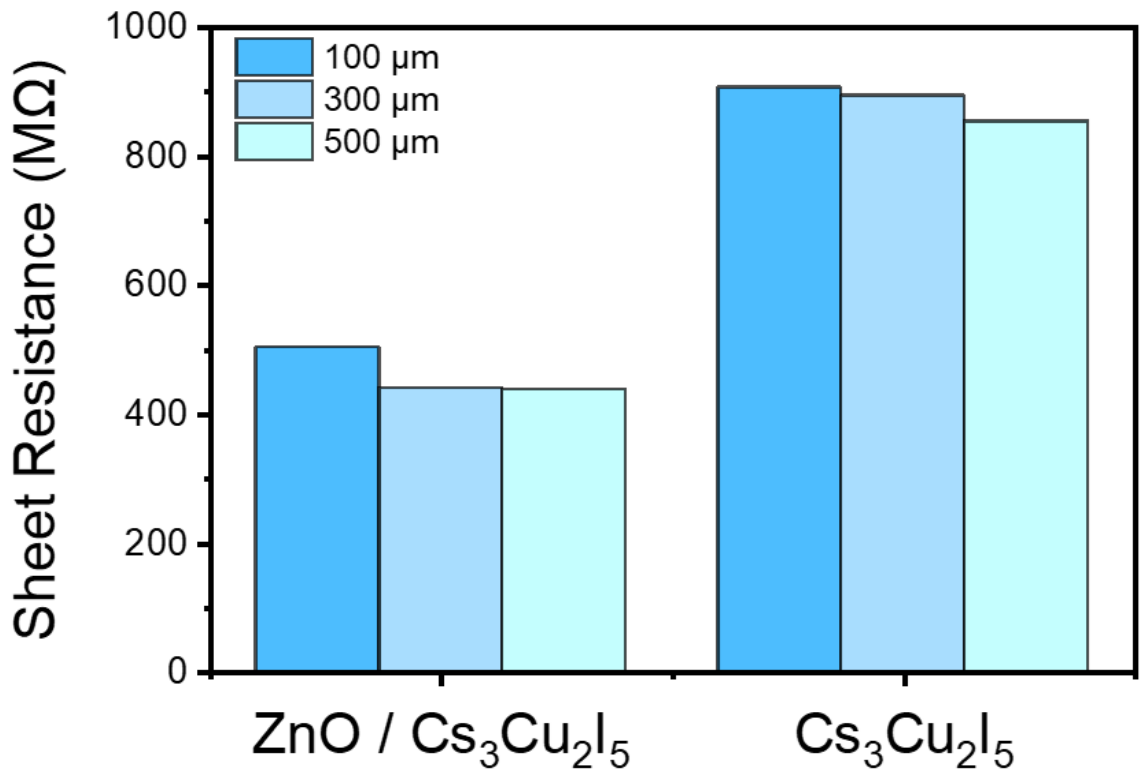

**Fig. 4.** Sheet resistance of $Cs_3Cu_2I_5$ and $ZnO/Cs_3Cu_2I_5$ heterostructures measured as a function of electrode distance.

**Fig. 4** presents a quantitative comparison of the sheet resistance (MΩ) for the pristine $Cs_3Cu_2I_5$ film and the $ZnO/Cs_3Cu_2I_5$ heterojunction structure as a function of electrode distance (100 μm, 300 μm, and 500 μm). Notably, these resistance values were obtained from current measurements conducted in the dark, reflecting the intrinsic electrical conductivity of the films before photoexcitation. This data directly demonstrates the effectiveness of the structural engineering strategy in improving lateral charge transport efficiency, which is a major challenge for $Cs_3Cu_2I_5$-based photodetectors.

The pristine $Cs_3Cu_2I_5$ film shows very high sheet resistance values reaching 900 MΩ. This high resistance is consistent with the material's inherently low electrical conductivity and the inefficient charge percolation network resulting from its isolated crystallite growth. In sharp contrast, the ZnO/Cs$_3$Cu$_2$I$_5$ heterojunction with the nanorippled ZnO layer exhibits a substantially reduced sheet resistance across all electrode distances. The resistance values for the heterojunction range from 400 MΩ to 500 MΩ, corresponding to an approximately 50% reduction compared to the pristine film. This pronounced decrease clearly indicates that the optimized ZnO nanoripple structure successfully provides new and continuous charge percolation paths connecting the otherwise isolated Cs$_3$Cu$_2$I$_5$ domains. These results confirm that the ZnO nanostructure plays a decisive role in overcoming the intrinsic electrical limitations of the 0D material and enabling enhanced lateral charge transport.

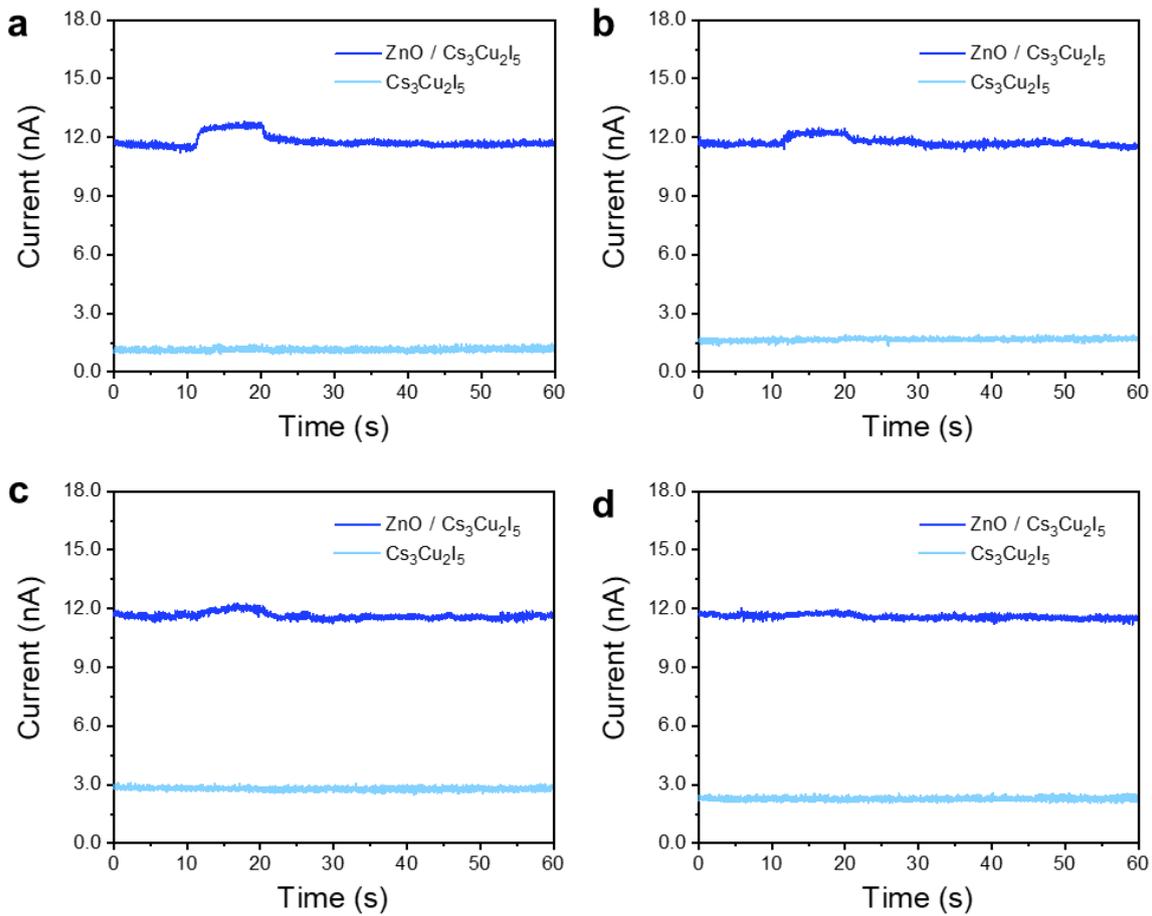

**Fig. 5.** Photocurrent generation by Cs$_3$Cu$_2$I$_5$ and ZnO/Cs$_3$Cu$_2$I$_5$ heterojunctions excited by **a** 390 nm, **b** 450 nm, **c** 515 nm, and **d** 660 nm.

**Fig. 5** compares the time-resolved photocurrent responses of the pristine $Cs_3Cu_2I_5$ film (light blue line) and the ZnO/ $Cs_3Cu_2I_5$ heterojunction (dark blue line) under excitation with four different monochromatic light sources: **Fig. 5a** 390 nm, **Fig. 5b** 450 nm, **Fig. 5b** 515 nm, and **Fig. 5d** 660 nm, respectively. A clear and consistent difference is observed across all measurements. The pristine $Cs_3Cu_2I_5$ film generates a very low photocurrent, remaining around 2.0–3.0 nA. This weak response aligns with the results shown in **Fig. 4**, confirming that the inherently low electrical conductivity of 0D material and inefficient lateral charge transport severely restrict its photodetection performance. In contrast, the ZnO $Cs_3Cu_2I_5$ heterojunction exhibits a markedly enhanced photocurrent under all excitation wavelengths, with stable current levels around 12–14 nA. This muti-fold improvement clearly demonstrates the effectiveness of the nanostructure-mediated charge transport engineering strategy. The introduction of the nanorippled ZnO layer establishes efficient percolation pathways that greatly enhance the collection of photogenerated carriers, resulting in a substantially amplified photocurrent response. This improvement is consistent across the visible and near-UV spectral regions, confirming the broad effectiveness of the optimized heterojunction design.

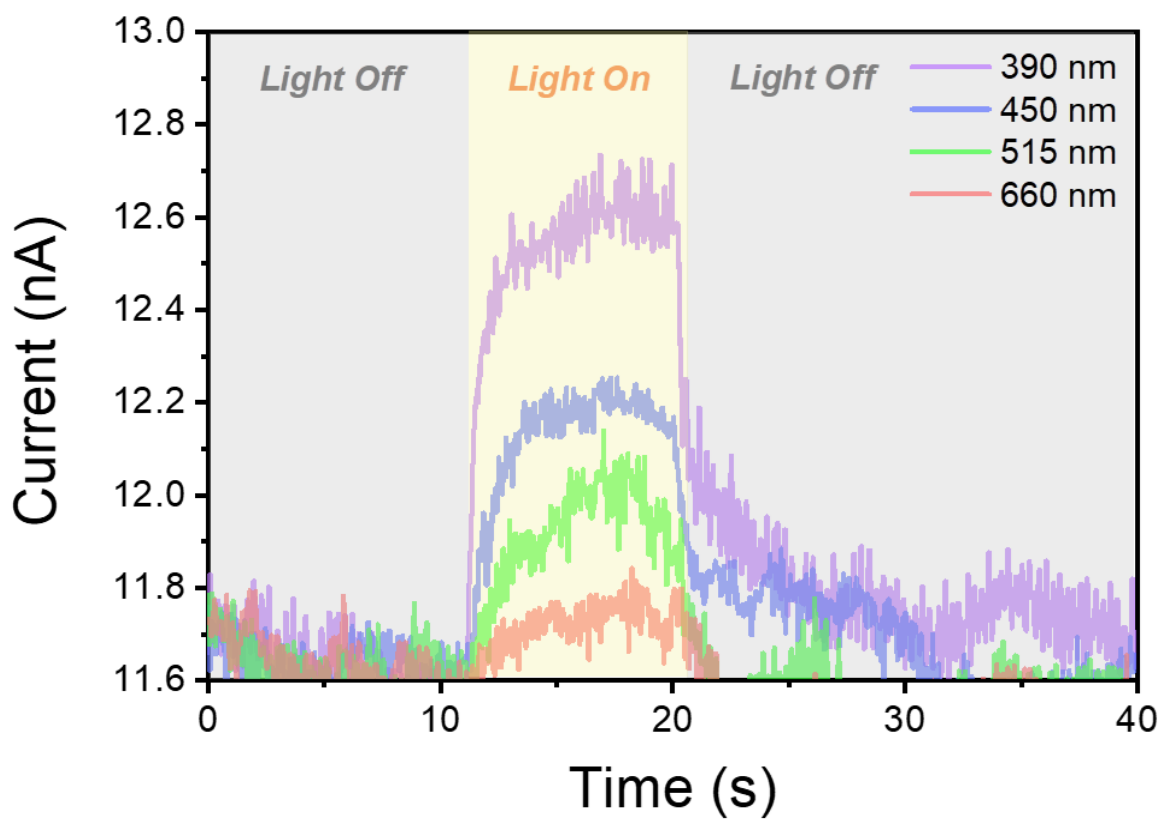

**Fig. 6.** Excitation wavelength-dependent photocurrent generated by ZnO/$Cs_3Cu_2I_5$ heterojunctions.

**Fig. 6** presents a unified plot of the time-resolved photocurrent response of the optimized ZnO/$Cs_3Cu_2I_5$ heterojunction (the structure proven superior in **Fig. 5**) under illumination by various monochromatic light sources (390 nm, 450 nm, 515 nm, and 660 nm). This figure clearly illustrates the device's spectral sensitivity and dynamic switching behavior between the "Light Off" (dark current) and "Light On" (photocurrent) states. Upon illumination, the device shows an immediate increase in current, demonstrating a fast photo-response. Notably, the photocurrent magnitude strongly depends on the excitation wavelength. The highest photocurrent is consistently observed under 390 nm excitation, where the current peaks above 12.6 nA. As the excitation wavelength increases toward the visible and red regions (450 nm →

515 nm → 660 nm), the photocurrent gradually decreases. This wavelength dependence directly reflects the absorption profile of the $Cs_3Cu_2I_5$ active layer, which, as a blue-emitting material, exhibits stronger absorption at shorter (higher-energy) wavelengths, resulting in the maximum response at 390 nm. The heterojunction maintains a clear and stable photo-response across the broad 390–660 nm range, confirming its capability for broadband photodetection. When the illumination is turned off, the current rapidly returns to its initial dark current level (approximately 11.7 nA). This swift recovery, together with the enhanced photocurrent, indicates that the integrated nano-rippled ZnO layer facilitates efficient charge extraction and suppresses carrier recombination which are the key factors for achieving fast and stable photodetector operation.

## 4. Conclusions

In summary, we demonstrated a novel strategy to overcome the intrinsic charge transport limitations of the promising yet poorly conductive 0D perovskite-inspired material, $Cs_3Cu_2I_5$. This was achieved by engineering a heterogeneous interface using an optimized ZnO nanoripple within a lateral photodetector (PD) architecture. Systematic control of the ZnO annealing temperature enabled the formation of an optimal interfacial topography that creates continuous percolation pathways between the otherwise isolated $Cs_3Cu_2I_5$ crystallites. This structural engineering led to a dramatic improvement in electrical performance, as evidenced by an approximately 50% reduction in sheet resistance for the $ZnO/Cs_3Cu_2I_5$ heterojunction compared with the pristine $Cs_3Cu_2I_5$ film. Moreover, the photoluminescence (PL) quenching observed in the heterojunction provided crucial optical evidence for the activation of efficient charge separation and extraction at the $ZnO/Cs_3Cu_2I_5$ interface. Ultimately, the photodetector exhibited significantly enhanced performance, showing a multifold increase in photocurrent response across the visible spectrum. The optimized device also displayed fast response times and distinct spectral selectivity, with the strongest response under 390 nm excitation. Overall, this work establishes nanoripple engineering as a powerful and practical approach to maximize charge collection efficiency which is emphasized in lateral devices based on "low conductivity 0D halide materials". Beyond conventional optoelectronics, this strategy holds strong potential for integrating stable, low-toxicity materials into high-sensitivity applications, in future Internet of Things (IoT) sensor networks and wearable electronics that require low power consumption and reliable environmental monitoring.

## 5. Acknowledgements

S.H.C. acknowledges that this research was supported by the Sungkyunkwan University and the BK21 FOUR (Graduate School Innovation) funded by the Ministry of Education (MOE, Korea) and National Research Foundation of Korea (NRF).

## 6. Funding

This work was supported by the National Research Foundation of Korea (NRF) grant funded by the Korea government (MSIT) (No. RS-2025-02217073, RS-2025-02302979) and the Regional Innovation System & Education (RISE) program through the (Chungbuk Regional Innovation System & Education Center), funded by the Ministry of Education (MOE) and the (Chungcheongbuk-do), Republic of Korea. (2025-RISE-11-014-03).

.